\documentclass{svjour2}                    
\smartqed  
\usepackage{mathptmx}      
\usepackage{bm,amsmath,amsfonts,color,graphicx}
\usepackage{textcomp,gensymb,units}
\usepackage[utf8]{inputenc}
\usepackage[pdftex]{hyperref}
\newcommand{\kt}{k_\mathrm{B}T}
\newcommand{\average}[1]{\left<{#1}\right>}

\renewcommand{\SS}{\mathsf{S}}
\journalname{Journal of Statistical Physics}
\begin{document}
\title{Einstein's Approach to Statistical Mechanics: 
The 1902--04 Papers}
\author{Luca Peliti \and Raúl Rechtman}
\institute{L. Peliti \at M. A. and~H.~Chooljan Member, Simons Center for Systems Biology,\\
Institute for Advanced Study, Einstein Drive, Princeton NJ 08540 (USA)\\
\href{mailto:luca@peliti.org}{\email{luca@peliti.org}}
\and
R. Rechtman \at Instituto de Energías Renovables, Universidad Nacional Autónoma de México,\\
Priv.\ Xochicalco S/N, Temixco, Morelos 62580 (México)\\
\href{mailto:rrs@ier.unam.mx}{\email{rrs@ier.unam.mx}}}

\date{Received: date / Accepted: date}

\maketitle

\begin{abstract}
We summarize the papers published by Einstein in the \textit{Annalen der Physik} in the years 1902--04 on the derivation of the properties of thermal equilibrium on the basis of the mechanical equations of motion and of the calculus of probabilities. We point out the line of thought that led Einstein to an especially economical foundation of the discipline, and to focus on fluctuations of the energy as a possible tool for establishing the validity of this foundation. We also sketch a comparison of Einstein's approach with that of Gibbs, suggesting that although they obtained similar results, they had different motivations and interpreted them in very different ways.
\end{abstract}

\section{Introduction}
\label{sec:intro}
By the end of June 1902, just after being accepted as Technical Assistant level III at the Federal Patent Office in Bern, Albert Einstein, 23, sent to the renowned journal \textsl{Annalen der Physik} a manuscript with the bold title ``Kinetic Theory of Thermal Equilibrium and of the Second Law of Thermodynamics''~\cite{Einstein02}.  In the introduction, he explains that he wishes to fill a gap in the foundations of the general theory of heat, ``for one has not yet succeeded in deriving the laws of thermal equilibrium and the second law of thermodynamics using only the equations of mechanics and the probability calculus''. He also announces ``an extension of the second law that is of importance for the application of thermodynamics''. Finally, he will provide ``the mathematical expression of the entropy from the standpoint of mechanics''. Einstein's papers and their translations are available on the Princeton University Press site~\cite{Princeton}. 

In the following two years Einstein followed this line of research publishing a paper each year~\cite{Einstein03,Einstein04}. The third one, entitled ``On the general molecular theory of heat'', submitted on  March 27, 1904, opened a new path, by tacitly extending the results obtained for a general mechanical system (with a large, but finite, number of degrees of freedom) to the case of black-body radiation. In pursuing this line of research Einstein found an unexpected result, that pointed at an inconsistency between the current understanding of the processes of light emission and absorption and the statistical approach. To resolve this inconsistency, in the first paper~\cite{Einstein05a} of his ``Annus Mirabilis'' 1905, Einstein renounced the detailed picture of light emission and adsorption provided by Maxwell's equations, maintaining his statistical approach, in particular the statistical interpretation of entropy. He introduced therefore the concept of light quanta, presented as a ``heuristic point of view''.

The importance of the 1902--04 papers on the molecular theory of heat in Einstein's intellectual development and in the advance of physics has been stressed by Kuhn~\cite[p.~171]{Kuhn}, when he states that
\begin{quotation}
What brought Einstein to the blackbody problem in 1904 and to Planck in 1906 was the coherent development of a research program begun in 1902, a program so nearly independent of Planck's that it would almost certainly have led to the blackbody law even if Planck had never lived.
\end{quotation}

In spite of their importance, the 1902--04 papers have received comparatively little attention. One of the reasons was the publication in 1902 of Gibbs' \textsl{Elementary Principles in Statistical Mechanics}.~\cite{Gibbs}  This book is considered, especially since the publication of the influential book by R.~C.~Tolman~\cite{Tolman}, as the founding text of the discipline. Einstein himself contributed to the neglect of the 1902-1904 papers. In his answer to Paul Hertz' criticism of his derivation of the second principle~\cite{Einstein11}, he says
\begin{quotation}
I only wish to add that the road taken by Gibbs in his book, which consists in one’s starting directly from the canonical ensemble, is in our opinion preferable to the road I took. If I had known Gibbs' book at that time, I would have not published these papers at all, but I would have limited myself to the treatment of a few points.
\end{quotation}
In his scientific autobiography~\cite[p.~47]{Einstein49} Einstein returned to this point, saying
\begin{quotation}
Not acquainted with the earlier investigations by Boltzmann and Gibbs, which had appeared earlier and actually exhausted the subject, I developed the statistical mechanics and molecular-kinetic theory of thermodynamics which was based on the former. My major aim in this was to find facts which would guarantee as much as possible the existence of atoms of definite size.
\end{quotation}

The last sentence of this quotation highlights the different attitude of Einstein with respect to Gibbs. Einstein aims at using the statistical approach to establish the reality of atoms, while Gibbs aims at a rational foundation of thermodynamics, and consequently focuses on the regularities which emerge in systems with many degrees of freedom. Einstein's papers contain a more direct and fundamental approach to the statistical mechanics of equilibrium, and could actually suggest a didactically effective path to the introduction of the fundamental ideas of the field. We shall therefore attempt to to ease their reading by summarizing them, pointing out in particular the differences between Einstein's and Gibbs' points of view. We shall not try to discuss all the detailed analyses of the papers which have appeared in the literature (beyond Kuhn's work~\cite{Kuhn}, one can also read \cite{Mehra75,BaraccaRechtman,Gearhart,Navarro,Uffink,Inaba}), but shall only refer to the more interesting observations. 
\section{Kinetic theory of thermal equilibrium and of the second principle of thermodynamics}
\label{sec:1902}
The first two papers~\cite{Einstein02,Einstein03} have a very similar structure. The second paper aims to widen the scope of the first, by attempting to consider ``general'' dynamical systems and irreversible processes. We shall follow the first paper, and we shall then briefly review the points in which the second paper differs. We adapt Einsteins discussion to modern notation.

Einstein begins by considering a general physical system as represented by a mechanical system with many coordinates $q=(q_{1},\ldots,q_{n})$ and the corresponding momenta $p=(p_{1},\ldots,p_{n})$, obeying the canonical equations of motion with a time-independent Hamiltonian that is the sum of a potential energy (function of the $q$'s alone) and of a kinetic energy that is a quadratic function of the $p$'s, whose coefficients are arbitrary functions of the $q$'s (and is implicitly supposed to be positive definite). Following Gibbs, we shall call the $p$'s and $q$'s collectively as the phase variables, and the space they span the phase space. Einstein then considers a very large number~$N$ of such systems, with the same Hamiltonian, whose energies $E$ lie between two very close values $\overline{E}$ and~$\overline{E}+\delta E$. He then looks for the stationary distribution of these systems in phase space.

Here Einstein introduces a strong mechanical hypothesis by assuming that, apart from the energy, there is no other function defined on the phase space that is constant in time.\footnote{This is the fundamental hypothesis linking the mechanical and the statistical aspects of the problem. It is probably inspired by the consideration of monocyclic systems, introduced by Helmholtz~\cite{Helmholtz} and discussed by Boltzmann in~\cite{BoltzmannMonocyclic}. Cf.~\cite{KleinMonocyclic} and~\cite{GallavottiErgodic}.} He argues that this condition is equivalent to the requirement that the stationary distribution of the systems in phase space depends only on the value of the energy. He proves indeed that if there are other functions $\phi(q,p)$ that are constants of the motion, the stationary distribution is not uniquely identified by the value of the energy, but does not attempt to prove the converse. He then shows that Liouville's theorem implies that the local density of systems in phase space is constant in time and therefore, by the mentioned hypothesis, must be a function of the energy alone. Since the energies of all $N$ systems are infinitely close to one another, this density must be uniform on the region of phase space defined by the corresponding value of the Hamiltonian. In this way Einstein has defined what is now called the microcanonical ensemble, i.e., the distribution in phase space which is uniform when the energy of the system lies between two closely lying values, and vanishes otherwise.

Einstein now turns to the consideration of thermal equilibrium between one system $\SS$ and one $\Sigma$ considerably larger.\footnote{Einstein actually considers two systems with the \textit{same} number of degrees of freedom, but where the energy contained in $\Sigma$ is considerably larger. Apparently the equipartition theorem, which he derives in \S~6 of the paper, led him to realize the awkwardness of this restriction, and he drops it in the second paper.} The second system acts as a thermal reservoir, and the first one as a thermometer. He assumes that the total energy $\mathcal{E}$ of the global system $\SS\cup\Sigma$ can be written as
\begin{equation}
\mathcal{E}=E+H,
\end{equation}
up to negligible terms, where $E$ pertains to $\SS$ and $H$ to~$\Sigma$. Let the phase variables of $\SS$ be denoted by $(p,q)$ and those of of $\Sigma$ by $(\pi,\chi)$. The question is now to find the distribution of the phase variables of $\SS$ when the energy of the global system lies between $\mathcal{E}_{0}$ and~$\mathcal{E}_{0}+\delta\mathcal{E}$, while the phase variables of~$\Sigma$ can take on any values. As pointed out by Uffink~\cite{Uffink}, this problem was considered several times by Boltzmann, who almost always solved it by taking an ideal gas for~$\Sigma$ and explicitly evaluating the resulting phase-space integral. Einstein instead introduces an elegant trick which leads directly to the desired result. Let us consider an infinitesimally small domain $g$ in the phase space of the global system $\SS\cup\Sigma$, with energy~$\mathcal{E}$ between $\mathcal{E}_{0}$ and~$\mathcal{E}_{0}+\delta\mathcal{E}$. Then the number $d N$ of systems of the ensemble which are found in $g$ is 
\begin{equation}
d N = A \int_{g}d p\,d q\; d\pi\, d\chi,
\end{equation}
where $A$ is a constant. Actually one can choose instead of $A$ any function of the total energy $\mathcal{E}$ which takes the value $A$ for $\mathcal{E}=\mathcal{E}_{0}$. Let us thus set\footnote{Einstein actually uses the notation $2h$ instead of $\beta$, which is now the traditional choice.}
\begin{equation}
A = A' \,e^{-\beta\, \mathcal{E}_{0}}=A'\,e^{-\beta\,E}e^{-\beta\,H },
\end{equation}
where $\beta$ is a constant. Thus the number $d N'$ of systems such that the phase variables of~$\SS$ lie in a region of volume $d p\;d q$ around the point $(p,q)$, while the variables of~$\Sigma$ can have any value, as long as $\mathcal{E}$ lies between $\mathcal{E}_{0}$ and $\mathcal{E}_{0}+\delta\mathcal{E}$, is given by
\begin{equation}\label{eq:integral}
d N' = A' e^{-\beta E}\,d p\, d q\int e^{-\beta H}\,d \pi\, d \chi,
\end{equation}
where the integral runs over all values of the phase variables of~$\Sigma$ such that the values of its Hamiltonian $H$ lie between $H_{0}$ and $H_{0}+\delta\mathcal{E}$, and
\begin{equation}
H_{0}=\mathcal{E}_{0}-E.
\end{equation}
The value of the constant $\beta$ can be fixed by requiring that the integral appearing on the right-hand side of~equation~(\ref{eq:integral}) be independent of~$E$. Indeed, once $\delta\mathcal{E}$ is fixed, the integral can be considered as a function $\Phi(H)$ of~$H$ alone. Thus, since $E\ll \mathcal{E}_{0}$, we have
\begin{equation}
\Phi(H_{0})=\Phi(\mathcal{E}_{0}-E)\simeq \Phi(\mathcal{E}_{0})-E\,\Phi'(\mathcal{E}_{0}),
\end{equation}
where $\Phi'$ is the derivative of~$\Phi$ with respect to its argument. Thus $\Phi'(\mathcal{E}_{0})=0$. We can write however
\begin{equation}
\Phi(H)=e^{-\beta H}\cdot \omega(H),
\end{equation}
where $\omega(H)=\int d\pi_{1}\cdots d\chi_{n}$, with the integral extended to the region in phase space such that the energy of $\Sigma$ lies between $H$ and $H+\delta\mathcal{E}$. The condition now reads
\begin{equation}
e^{-\beta \mathcal{E}_{0}}\omega(\mathcal{E}_{0})\left[-\beta +\frac{\omega'(\mathcal{E}_{0})}{\omega(\mathcal{E}_{0})}\right]=0,
\end{equation}
where $\omega'$ is the derivative of $\omega$ with respect to is argument. We therefore obtain the required condition for $\beta$ in the form
\begin{equation}
\beta = \frac{\omega'(\mathcal{E}_{0})}{\omega(\mathcal{E}_{0})}.
\end{equation}

Einstein now turns to show that the quantity $\beta$ is always positive. He first derives a lemma, by considering a general (positive definite) quadratic function $\varphi(x_{1},\ldots,x_{n})$ of $n$ variables (where $n$ is large enough), and defining the function $z(y)$ by the integral
\begin{equation}
z(y)=\int d x_{1}\cdots d x_{n},
\end{equation}
where the integral is extended to all points for which $\varphi$ lies between $y$ and $y+\Delta$, where $\Delta$ is fixed. He then easily shows that, for $n\ge 3$, $z(y)$ is an increasing function of $y$. Let us now denote by $\Gamma(H)$ the phase space available to the larger system~$\Sigma$ when the values of its Hamiltonian lie between $H$ and $H+\delta\mathcal{E}$. The Hamiltonian of $\Sigma$ is given by the sum of the potential energy, that depends only on the coordinates, and of the kinetic energy, which is a quadratic form in the momenta, whose coefficients depend only on the coordinates. Let $H_{0}$ and $H_{1}$ be two values of $H$, with $H_{1}>H_{0}$, and let $\Gamma(H_{0})$ and $\Gamma(H_{1})$ be the corresponding available space regions. Let $Q(H_{0})$ be the region of \textit{coordinate space} such that the potential energy of the system is smaller than $H_{0}$. Thus if the point $(\pi,\chi)$ belongs to $\Gamma(H_{0})$, the point $(\chi)$ belongs to $Q(H_{0})$. Within $\Gamma(H_{1})$ let us identify the region $\Gamma'(H_{1})$ where the coordinates $\chi$ belong to $Q(H_{0})$.  Thus, for each such values of the coordinates, since the total energy is larger than $H_{0}$, the kinetic energy must be larger. Therefore, by the lemma on the monotonic increase of~$z(y)$ with $y$, for each such point in coordinate space, the volume available to the momenta is larger for $\Gamma'(H_{1})$ than for $\Gamma(H_{0})$. Integrating over the coordinates we obtain that the volume of $\Gamma'(H_{1})$ must be larger than that of $\Gamma(H_{0})$. Since the volume of the region of $\Gamma(H_{1})$ that does not belong to $\Gamma'(H_{1})$ cannot be negative, the volume of $\Gamma(H_{1})$ must be larger than that of $\Gamma(H_{0})$, i.e., the function $\omega(H)$ increases with $H$, and $\beta$ given by the above expression must be positive.

Now, Einstein derives what is now known as the zero-th law of thermodynamics. Since $\beta$ depends only on the state of~$\Sigma$, but determines the distribution of $\SS$ in state space, independently on how $\Sigma$ and $\SS$ interact, it follows that if a given system $\Sigma$ interacts with two small system $\SS$ and~$\SS'$ and is in equilibrium with them, $\SS$ and~$\SS'$ must have the same value of~$\beta$. In particular, if $\SS$ and $\SS'$ are mechanically identical, the average value of any arbitrary observable function $A(p,q)$ must be equal in $\SS$ and~$\SS'$. Einstein then calls $\SS$ and~$\SS'$ thermometers, $\beta$ the temperature function and the average of~$A$ the temperature measure. Then Einstein goes on to prove the converse result, namely that if two systems that have the same values of~$\beta$ are put in contact, they will be in thermal equilibrium. He considers two systems, $\Sigma_{1}$ and~$\Sigma_{2}$, weakly interacting. Let each of them be in contact with an (infinitesimally) small thermometer $\SS_{1}$ and~$\SS_{2}$. The temperature measures $A_{1}$ and $A_{2}$ in each thermometer will be the same, since we are in fact dealing with a single interacting system in thermal equilibrium, and therefore also the corresponding temperature functions $\beta_{1}$ and~$\beta_{2}$ will be equal. Let the interaction terms between $\Sigma_{1}$ and~$\Sigma_{2}$ be slowly brought to zero. Then the readings of the thermometers will remain equal, but now the reading of $\SS_{1}$ deals only with $\Sigma_{1}$ and that of~$\SS_{2}$ only with $\Sigma_{2}$. The process is reversible, since we are dealing with a sequence of thermal equilibrium states. Thus, by reversing it, we obtain the required result. As an immediate consequence, we obtain that if $\Sigma_{1}$ and $\Sigma_{2}$ are in thermal equilibrium, and so are $\Sigma_{2}$ and~$\Sigma_{3}$, then $\Sigma_{1}$ and~$\Sigma_{3}$ are in thermal equilibrium, since they share the same value of~$\beta$. Einstein concludes this section with the intriguing remark:
\begin{quotation}
I would like to note here that until now we have made use of the assumption that our systems are mechanical only inasmuch as we have applied Liouville's theorem and the energy principle. Probably the basic laws of the theory of heat can be developed for systems that are defined in a much more general way. We will not attempt to do this here, but will rely on the equations of mechanics. We will not deal here with the important question as to how far the train of thought can be separated from the model employed and generalized.
\end{quotation}
Uffink~\cite{Uffink} has remarked that ``this quote indicates (with hindsight) a remarkable underestimation of the logical dependence of [Einstein's] approach on the ergodic hypothesis.'' But the passage shows, as also stressed by Uffink, that already in 1902 Einstein was considering the need to extend the statistical approach beyond its application to mechanical systems, no matter how general they can be conceived. 

A simple calculation allows Einstein to derive the equipartition theorem in the following form. Let the kinetic energy of a system be represented by a quadratic expression of the form
\begin{equation}
K=\frac{1}{2}\left(\alpha_{1}p_{1}^{2}+\cdots+\alpha_{n}p_{n}^{2}\right),
\end{equation}
where $\alpha_{i}$, $i=1,\ldots,n$, are positive constants or functions of the coordinates~$q$. This form can always be reached from a general quadratic expression by a suitable canonical transformation. The $p$ variables had been denoted as ``momentoids'' by Boltzmann. Then the average of~$K$ at equilibrium is given by
\begin{equation}
\average{K}=\frac{n}{2\beta}.
\end{equation}
In particular, this result implies that the kinetic energy of a single molecule in an ideal gas is equal to $3/(2\beta)$ on average. Kinetic theory teaches us that this quantity is proportional to the product of the pressure and the volume per particle in an ideal gas. Since this is proportional to the absolute temperature~$T$, we obtain
\begin{equation}
\frac{1}{\beta}=k_{\mathrm{B}} T=\frac{\omega(H)}{\omega'(H)},
\end{equation}
where $k_{\mathrm{B}}$ is Boltzmann's constant and~$\omega(H)$ is the volume of phase space contained by the equal-energy surfaces of~$\Sigma$ corresponding to the values $H$ and $H+\delta\mathcal{E}$.

Having found the relation between~$\beta$ and the temperature, Einstein proceeds to the derivation of the second law of thermodynamics, which he here limits to the statement of the integrability of heat divided by the absolute temperature. He switches back to a Lagrangian setting, describing the system's state by the coordinates~$q$ and their time derivatives~$\dot{q}$, and introduces externally applied forces. These forces are split into ones derived from a potential depending on the $q$'s, and others that allow for heat transfer. The first ones are assumed to vary slowly with time, while the second ones change very rapidly. The infinitesimal heat $\delta Q$ is defined as the work of the second type of forces. Then a reversible transformation is one in which the system is led from an equilibrium state with given values of~$\beta$ and of the volume~$V$ to one with the values $\beta+\delta\beta$ and~$V+\delta V$. Here Einstein tacitly assumes that the time average of the relevant quantities in a slow transformation can be obtained by averaging the same quantity over the distribution of the $N$ systems in phase space. He thus finds that
\begin{equation}
\frac{\delta Q}{T}=d\left(\frac{\average{E}-F}{T}\right),
\end{equation}
where $\average{E}$ is the average total energy of the system, and $F$ is a constant introduced so that the distribution $P(p, q)= e^{\beta(E(p,q)-F)}$ is normalized. Einstein remarks that this expression contains the total energy, and is independent of its splitting into kinetic and potential terms.\footnote{This will be the starting point of his 1903~paper.} One can readily integrate this expression, obtaining an explicit form of the entropy~$S$:
\begin{equation}
S =\frac{\average{E}-F}{T}=\frac{\average{E}}{T}+k_{\mathrm{B}}\log\int  e^{-\beta E(p,q)}\,d p\,d q+\text{const.}
\end{equation}

Now Einstein states the announced generalization of the second principle. It is worth quoting this short paragraph in its entirety. Einstein denotes by $V_{a}$ the potential of the conservative forces performing the reversible transformation. He then states
\begin{quotation}
No assumptions had to be made about the nature of the forces that correspond to the potential $V_{a}$ [the conservative ones], not even that such forces occur in nature. Thus, the mechanical theory of heat requires that we arrive at correct results if we apply Carnot's principle to ideal processes, which can be produced from the observed processes by introducing arbitrarily chosen $V_{a}$'s. Of course, the results obtained from the theoretical consideration of those processes can have real meaning only when the ideal auxiliary forces $V_{a}$ no longer appear in them.
\end{quotation}
Thus the strategy which led to the establishment of the Einstein relation in Brownian motion, in the 1905 paper, is already sketched in this one.
\section{A theory of the foundations of thermodynamics}
\label{sec:1903}
In his 1903 memoir, entitled ``A theory of the foundations of thermodynamics''~\cite{Einstein03}, Einstein asks whether kinetic theory is essential for the derivation of the postulates of thermal equilibrium and of the entropy concept, or whether ``assumptions of a more general nature'' could be sufficient. He goes on therefore to consider a general dynamical system whose state is identified by a collection $p$ of variables $p=(p_{1},\ldots,p_{n})$, which correspond to both coordinates and momenta for a mechanical system, and evolve by a general system of equations of motion of the kind
\begin{equation}\label{eq:eqmoto}
\frac{d p_{i}}{d t}=\varphi_{i}(p_{1},\ldots,p_{n});\qquad i=1,\ldots,n.
\end{equation}
Assuming that the system allows for a unique integral of motion, the energy $E(p)$, he then introduces the equilibrium postulate, according to which a ``physical system'' eventually reaches a time-independent macroscopic state, in which any ``perceptible quantity'' assumes a time-independent value. Einstein then looks for the stationary distribution of a collection of~$N$ systems, with $N$ large. Each system evolves according to equations~(\ref{eq:eqmoto}) and has an energy between $E$ and~$E+\delta E$. He claims that the equilibrium postulate, along with the absence of any integral of motion beyond the energy, implies the existence of a well-defined probability distribution in $p$-space. Einstein's argument reads
\begin{quotation}
Starting at an arbitrary point of time and throughout time~$\mathcal{T}$, we consider a physical system which is represented by the equations~(\ref{eq:eqmoto}) and has the energy~$E$. If we imagine having chosen some arbitrary region~$\Gamma$ of the state variables~$p_{1}\ldots p_{n}$, then at a given instant of time $\mathcal{T}$ the values of the variables $p_{1}\ldots p_{n}$ will lie within the chosen region~$\Gamma$ or outside it; hence, during a fraction of the time~$\mathcal{T}$, which we will call~$\tau$, they will lie in the chosen region~$\Gamma$. Our condition then reads as follows: If the $p_{1}\ldots p_{n}$ are state variables of a physical system, i.e.,  of a system that assumes a stationary state, then for each region~$\Gamma$ the quantity $\tau/\mathcal{T}$ has a definite limiting value for $\mathcal{T}=\infty$. For each infinitesimally small region this value is infinitesimally small.
\end{quotation}
Thus the stationary distribution is identified by a function $\epsilon(p_{1},\ldots, p_{n})$ such that the number $d N$ of systems which at any given instant in time are found in the infinitesimal region $g$ located around $(p_{1},\ldots,p_{n})$ is given by
\begin{equation}
d N = \epsilon(p_{1},\ldots,p_{n})\,d p_{1}\cdots d p_{n}.
\end{equation}
If this is true at a given instant $t$, then at a close instant $t+d t$ one has
\begin{equation}
d N_{t+d t}=d N_{t}-\left(\sum_{\nu=1}^{n}\frac{\partial(\epsilon \varphi_{\nu})}{\partial p_{\nu}}\right)d p_{1}\cdots d p_{n}.
\end{equation}
Since $d N_{t+d t}=d N_{t}$, by the stationarity of the distribution, one must have
\begin{equation}
\sum_{\nu=1}^{n}\frac{\partial(\epsilon \varphi_{\nu})}{\partial p_{\nu}}=0.
\end{equation}
Then
\begin{equation}\label{eq:delogeps}
-\sum_{\nu=1}^{n}\frac{\partial \varphi_{\nu}}{\partial p_{\nu}}=\sum_{\nu=1}^{n}\frac{\partial\log\epsilon}{\partial p_{\nu}}\varphi_{\nu}=\frac{d\log\epsilon}{d t}.
\end{equation}
The solution of equation~(\ref{eq:delogeps}) is
\begin{equation}
\epsilon=\exp\left[-\int d t\;\sum_{\nu=1}^{n}\frac{\partial \varphi_{\nu}}{\partial p_{\nu}}+\psi(E)\right],
\end{equation}
where $\psi(E)$ is a time-independent integration constant that, by the previous hypotheses, can only depend on the~$p$'s via the energy $E$. One thus obtains
\begin{equation}
\epsilon=\text{const.}\times \exp\left[-\int dt\;\sum_{\nu=1}^{n}\frac{\partial \varphi_{\nu}}{\partial p_{\nu}}\right]=\text{const.}\; e^{-m},
\end{equation}
where $m$ is given by
\begin{equation}
m = \int dt\;\sum_{\nu=1}^{n}\frac{\partial \varphi_{\nu}}{\partial p_{\nu}}.
\end{equation}
Einstein now assumes that it is possible to introduce new state variables, denoted by $\pi_{1},\ldots,\pi_{n}$, such that the factor $e^{-m}$ is cancelled by the Jacobian of the transformation. With this transformation, one obtains a uniform stationary distribution in phase space. However it is clear that this transformation cannot be performed unless $m$ is time-independent, which implies $d(\log\epsilon)/d t=0$ throughout, i.e., a form of Liouville's theorem. The oversight was realized by Einstein in March~1903, as witnessed by a letter to Michele Besso,~\cite[Vol.~5, Doc.~7]{CPAE} quoted by Uffink~\cite{Uffink}:
\begin{quotation}
If you look at my paper more closely, you will find that the assumption of the energy principle \&\ of the fundamental atomistic idea alone does not suffice for an explanation of the second law; instead, coordinates $p$ must exist for the representation of things, such that for every conceivable total system $\sum \partial\phi_{\nu}/\partial p_{\nu}=0$. [\dots] If that is true, then the entire generalization attained in my last paper consists in the elimination of the concept of force as well as in the fact that $E$ can possess an arbitrary form (not completely)?
\end{quotation}

The sections that immediately follow, on the distribution of a system in contact with a reservoir, on the absolute temperature and thermal equilibrium, and on the definition of ``infinitely slow'' (quasistationary) processes, are not fundamentally different from the corresponding sections of the 1902~memoir. The derivation of the mechanical expression of the entropy is however slightly different, in particular because the possibility of resorting to the Lagrangian formulation is no longer available. Einstein considers a situation in which the functions $\varphi_{\nu}$ which appear on the right-hand side of the equations~(\ref{eq:eqmoto}) depend not only on the coordinates $p_{\nu}$, but also on some parameters $\lambda$. He then considers an infinitely-slow  infinitesimal transformation, subdividing it into an \textit{isopycnic process}, in which the $\lambda$'s are kept constant, but the system is put in thermal contact with a system at a different temperature, and an \textit{adiabatic process}, in which the system is isolated, but the $\lambda$'s are allowed to vary. The energy change $dE$ is given in general by
\begin{equation}\label{eq:dE}
dE=\sum\frac{\partial E}{\partial \lambda}d\lambda+\sum_{\nu}\frac{\partial E}{\partial p_{\nu}}dp_{\nu}.
\end{equation}
In an isopycnic process the first term on the right-hand side of this equation vanishes, but the second term can be different from zero, since the equations of motion~(\ref{eq:eqmoto}), which conserve~$E$, do not hold when the system is not isolated. In an adiabatic process, on the other hand, the second term vanishes, since the equations of motion~(\ref{eq:eqmoto}) satisfy energy conservation, but at the same time one has $dQ=0$. One can therefore write in general
\begin{equation}\label{eq:heat}
dQ=\sum_{\nu}\frac{\partial E}{\partial p_{\nu}}\,dp_{\nu}.
\end{equation}
Therefore, in the expression for the change of energy in an infinitely slow process given in equation~(\ref{eq:dE}), one can identify the second term in the right-hand side with the infinitesimal heat exchange $dQ$, and the first one, accordingly, with the infinitesimal work. Einstein has thus obtained a mechanical expression of the first principle of thermodynamics.

Let us now denote by $W(p_{1},\ldots,p_{n})$ the probability distribution in phase space of the system when it is in equilibrium with an external body with a temperature function given by $\beta$. As derived by Einstein in \S~3 of the paper, along the lines of its 1902 paper, it is given by
\begin{equation}
dW=e^{c-\beta E}\,dp_{1}\cdots dp_{n},
\end{equation}
where the constant~$c$ is defined by the normalization condition
\begin{equation}
\int dW=\int e^{c-\beta E}\,dp_{1}\cdots dp_{n}=1.
\end{equation}
Let us assume that after the transformation, the system is in equilibrium with a body with temperature function $\beta+d\beta$, while the parameters $\lambda$ assume the values $\lambda+d\lambda$. Then the normalization condition assumes the form
\begin{equation}
\int \exp\left[c+dc -(\beta+d\beta)\left(E+\sum\frac{\partial E}{\partial \lambda}d\lambda\right)\right]\,dp_{1}\cdots dp_{n}=1.
\end{equation}
One thus obtains, to first order,
\begin{equation}
\begin{split}
&\int \left(dc - E\,d\beta-\beta\sum\frac{\partial E}{\partial \lambda}d\lambda\right)\, e^{c-\beta E}\,dp_{1}\cdots dp_{n}=0.
\end{split}
\end{equation}
Einstein now argues that the expression in parentheses can be considered as a constant, ``because the system's energy $E$ never differs markedly from a fixed average before and after the process'', and thus obtains
\begin{equation}\label{eq:2nd}
dc - E\,d\beta-\beta\sum\frac{\partial E}{\partial \lambda}d\lambda=0.
\end{equation}
Since
\begin{equation}
E\,d\beta+\beta\sum\frac{\partial E}{\partial\lambda}d\lambda=d\left(\beta E\right)-\beta\sum_{\nu}\frac{\partial E}{\partial p_{\nu}}dp_{\nu}=d\left(\beta E\right)-\beta\,dQ,
\end{equation}
where equation~(\ref{eq:heat}) has been substituted, Einstein obtains the relation
\begin{equation}
\beta\,dQ=d(\beta E-c),
\end{equation}
and thus, since $1/\beta=\kt$,
\begin{equation}\label{eq:entropydiff}
\frac{dQ}{T}=d\left(\frac{E}{T}-k_{\mathrm{B}}c\right)=dS,
\end{equation}
from which he obtains the expression of the entropy
\begin{equation}\label{eq:entropy}
S=\frac{E}{T}-k_{\mathrm{B}}c=\frac{E}{T}+k_{\mathrm{B}}\log \int e^{-E/\kt}\,dp_{1}\cdots dp_{n}.
\end{equation}

It is interesting to remark that in the 1902 paper Einstein had derived a similar expression of the heat exchanged $dQ$ involving the \textit{average} values of the kinetic and potential energies, while here Einstein states that the values of the energy $E$ which matter are not very different from their mean value. This assumption is unnecessary because the relation (\ref{eq:2nd}) holds if $E$ is understood as the mean value of the energy, which is enough to reach Einstein's goals. Moreover, Einstein has not yet derived this property of the energy distribution. We shall see that this assumption also leads Einstein to a quite dubious result in the next discussion, where he attempts to establish the property of entropy increase. In our opinion, Einstein later reconsidered this argument and was led therefore to investigate the fluctuations of energy, which he discusses in his next paper. 

Einstein now attempts to prove that the entropy does not decrease in transformations involving an adiabatically isolated system. He goes on from the probability distribution of a single system in its phase space, when the value of its energy is fixed, to the distribution of a collection of a very large number~$N$ of such systems with the same value of the energy. Dividing the phase space in $\ell$ regions $g_{i}$, $i=1,\ldots,\ell$ of equal volume, Einstein looks for the probability of that $n_{1}$ systems fall in $g_{1}$, \dots, $n_{\ell}$ systems fall in $g_{\ell}$. The result is obviously
\begin{equation}\label{eq:logW}
W=\left(\frac{1}{\ell}\right)^{N}\frac{N!}{n_{1}!\cdots n_{\ell}!}.
\end{equation}
One then has, by Stirling's formula,
\begin{equation}\label{eq:stirling}
\begin{split}
\log W &=\text{const.}-\sum_{i} n_{i} \log n_{i} \simeq \text{const.}-\int \rho \log \rho\;dp_{1}\cdots dp_{n},
\end{split}
\end{equation}
where $\rho$ is the density of systems in the $p$-space, when $\ell\to\infty$. It would have been a simple step to connect explicitly this expression to the entropy by means of Boltzmann's formula, but Einstein does not do it. He instead uses it first to show that this expression reaches a maximum when $\rho$ is constant on the whole region of phase space in which the energy has the assigned value. He then argues that if the density $\rho$ differs noticeably from a constant (for states of a given value of the energy), it will be possible to find distributions with a larger value of~$W$. In this case, if we follow in time the ensemble, the distribution will change with time, and since ``we will have to assume that always more probable distributions will follow upon improbable ones, i.e., that $W$ increases until the distribution of states has become constant and $W$ a maximum''. Thus, if the distribution changes from $\rho$ to $\rho'$ as time goes by, and the probability correspondingly increases from $W$ to $W'$, the integral on the right-hand side of equation~(\ref{eq:stirling}) decreases. He then argues that if the values of $\log \rho$ (when $\rho$ does not essentially vanish) are close to uniform, and the probability increases, one obtains the  relation
\begin{equation}\label{eq:ineq}
-\log \rho'\ge -\log \rho.
\end{equation}
This equation cannot be true without qualification, due to the normalization condition, and is however unnecessary for Einstein's argument in the immediately following section. See, e.g., the discussion in~\cite[\S2.2]{Uffink}. This is probably one of the points which led Einstein, in retrospect, to reconsider the assumption that the values of the energy which have non-vanishing probability are close to constant, and to evaluate the energy fluctuations.

Einstein then takes advantage of this result to obtain the law of entropy increase in the following way. He considers a finite number of systems $\sigma_{1},\ldots,\sigma_{\nu},\ldots$, that together form an isolated system with state variables $p^{(1)}_{1},\ldots,p^{(1)}_{n_{1}},\ldots,$\break $p_{1}^{(\nu)},\ldots,p_{n_{\nu}}^{(\nu)},\ldots$, such that $n=\sum_{\nu}n_{\nu}$. System $\sigma_{\nu}$ is initially in equilibrium at a temperature $T_{\nu}=1/k_{\mathrm{B}}\beta_{\nu}$, and is therefore described by the distribution
\begin{equation}
dw_{\nu}=e^{c_{\nu}-\beta_{\nu}E_{\nu}}\,dp_{1}^{(\nu)}\cdots dp_{n_{\nu}}^{(\nu)}.
\end{equation}
Then the distribution of the global system is given by
\begin{equation}
dw=\prod_{\nu}dw_{\nu}=e^{\sum(c_{\nu}-\beta_{\nu}E_{\nu})}\,dp_{1}\cdots dp_{n}.
\end{equation}
Let us assume that the systems are now allowed to interact among themselves, and that at the end of the process a new equilibrium is reached, characterized by the temperature parameters $\beta'_{\nu}$, etc. We then have, at the end of the process,
\begin{equation}
dw'=\prod_{\nu}dw'_{\nu}=e^{\sum(c'_{\nu}-\beta'_{\nu}E'_{\nu})}\,dp_{1}\cdots dp_{n}.
\end{equation}
Einstein now introduces an ensemble of a very large number~$N$ of global systems~$\Sigma$ to argue that, since $W$ always increases, the distributions
\begin{subequations}\label{eq:distrib}
\begin{align}
\rho &= N \,e^{\sum(c_{\nu}-\beta E_{\nu})};\\
\rho' &= N \,e^{\sum(c'_{\nu}-\beta' E'_{\nu})};
\end{align}
\end{subequations}
satisfy equation~(\ref{eq:ineq}), i.e.,
\begin{equation}\label{eq:deltaS}
\sum\left(c'_{\nu}-\beta'_{\nu} E'_{\nu}\right)\ge \sum(c_{\nu}-\beta_{\nu}E_{\nu}).
\end{equation}
But this implies, by equation~(\ref{eq:entropy}),
\begin{equation}
\sum S'_{\nu}\ge \sum S_{\nu}.
\end{equation}
Again, the detour by equation~(\ref{eq:ineq}) is disputable and unnecessary. Indeed, it is sufficient to use equation~(\ref{eq:logW}) to obtain equation~(\ref{eq:deltaS}) where $E$ is now taken as the mean value of the energy, and the result would follow. The observations made after equation (\ref{eq:2nd}) also apply here. However, the main weakness of the argument lies in the \textit{petitio principii} that the probability $W$ of the ensemble distribution should always increase. This objection was raised by Paul Hertz in 1910~\cite{Hertz}, and Einstein soon acknowledged~\cite{Einstein11} that the objection was ``fully founded''.

In the closing section of this paper Einstein applies these results to a simple description of a thermal engine connected in turn to several heat reservoirs to derive the second principle in the form of Clausius.
\section{On the general molecular theory of heat}
\label{sec:04}
A change of pace is easily noticed already in the first lines of the 1904 paper, entitled ``On the general molecular theory of heat.''\cite{Einstein04} Here he refers to his previous papers, in which he had spoken of the ``kinetic theory of heat'' as laying the foundations of thermodynamics, by the less specific expression of ``molecular theory of heat''. The paper contains several results worth mentioning, as announced at the end of the introduction
\begin{quotation}
First, I derive an expression for the entropy of a system, which is completely analogous to the expression found by Boltzmann for ideal gases and assumed by Planck in his theory of radiation. Then I give a simple derivation of the second law. After that I examine the meaning of a universal constant, which plays an important role in the general molecular theory of heat. I conclude with an application of the theory to black-body radiation, which yields a most interesting relationship between the above-mentioned universal constant, which is determined by the magnitudes of the elementary quanta of matter and electricity, and the order of magnitude of the radiation wave-lengths, without recourse to special hypotheses.
\end{quotation}
These results are obtained as independent developments of the theory reported in the previous two papers. In the previous papers he had derived the canonical expression of entropy, namely
\begin{equation}
S=\frac{E}{T}+k_{\mathrm{B}}\int  e^{-E/\kt}\,dp_{1}\cdots dp_{n},
\end{equation}
where $(p_{1},\ldots,p_{n})$ are the general state variables of the system, and $E$ is the value of the internal energy. In \S~1 of this paper Einstein derives the expression we now call microcanonical, which is related to the density of states of energy $E$, $\omega(E)$, by the relation
\begin{equation}
S=k_{\mathrm{B}}\log[\omega(E)].
\end{equation}
He obtains this result by integrating the relation between the temperature and $\omega(E)$ previously derived:
\begin{equation}
\frac{1}{k_{\mathrm{B}}T}=\frac{\omega'(E)}{\omega(E)},
\end{equation}
where one assumes that the system's energy lies between $E$ and~$E+\delta E$. Note, however, that in the previous papers $\omega(E)$ was the energy density of the thermal reservoir, while this relation is tacitly applied here to the energy density of the system. Interestingly, in this paper Einstein defines for the first time the density of states $\omega(E)$ in the now customary way, by
\begin{equation}
\omega(E)\,\delta E=\int_{E}^{E+\delta E}d p_{1}\cdots d p_{n},
\end{equation}
while in the previous papers he kept including the $\delta E$ factor in its definition.

The ``derivation'' of the second law in~\S~2 suffers again, as in the 1903 paper, from the \textit{petitio principii} of the assumption that more improbable states never follow more probable ones. The calculation is now simpler, but the result is also more restricted. First Einstein formulates the zero-th laws law of thermodynamics by assuming that if a system is in contact with an environment at temperature $T_{0}$ it acquires the temperature $T_{0}$ and keeps it from then on. However, according to the molecular theory of heat, this is not absolutely true, but true only with some approximation. In particular the probability $W\,\delta E$ that the energy of such a system has a value lying between $E$ and $E+\delta E$ at an arbitrary point in time is given by
\begin{equation}
W\,\delta E=C \, e^{-E/\kt_{0}}\,\omega(E) \,\delta E,
\end{equation}
where $C$ is a constant. Einstein argues that this distribution is very sharply peaked and that, because of the previous result, it can also be written in the form
\begin{equation}
W\,\delta E=C\, \exp\left[\frac{1}{k_{\mathrm{B}}}\left(S-\frac{E}{T_{0}}\right)\right]\,\delta E,
\end{equation}
where $S=S(E)$ is the value of the entropy pertaining to the value $E$ of the internal energy. Note that here again the property of the distribution of being sharply peaked is not needed, and anyway has not yet been derived. More interestingly, as far as we know, this is the first statement of Einstein's principle of fluctuations, which relates the probability of an energy fluctuation in a thermodynamic system to the difference in the expression $\mathcal{F}(E,T)=E-TS(E)$, which is now known as the availability. Now Einstein considers a system made of several such subsystems, all in contact with a large similar system at the temperature $T_{0}$. The probability $\mathfrak{W}$ of a given distribution $(E_{1},\ldots,E_{\ell})$ of the energy among these subsystems is given by
\begin{equation}
\mathfrak{W}\propto \exp\left[\frac{1}{k_{\mathrm{B}}}\left(\sum_{i=1}^{\ell}S_{i}-\frac{1}{T_{0}}\sum_{i=1}^{\ell}E_{i}\right)\right].
\end{equation}
Let the reservoirs exchange energy, maybe by the assistance of cyclic machines, reaching an energy distribution $(E'_{1},\ldots,E'_{\ell})$. The corresponding probability is given by
\begin{equation}
\mathfrak{W}'\propto \exp\left[\frac{1}{k_{\mathrm{B}}}\left(\sum_{i=1}^{\ell}S'_{i}-\frac{1}{T_{0}}\sum_{i=1}^{\ell}E'_{i}\right)\right].
\end{equation}
Assuming again that less probable states are followed by more probable ones, one must have
\begin{equation}
\mathfrak{W}'\ge \mathfrak{W}.
\end{equation}
Since $\sum_{i}E_{i}$ is conserved, this equation implies
\begin{equation}
\sum_{i=1}^{\ell}S'_{i}\ge \sum_{i=1}^{\ell}S_{i}.
\end{equation}
It is hard for us to make sense of this derivation. The results seems restricted to systems in contact with a reservoir with a given temperature~$T_{0}$, and therefore it is by no means general. In particular the inequality among the~$\mathfrak{W}$'s cannot be \textit{absolutely} satisfied without violating the normalization of probabilities, just as in the case of equation~(\ref{eq:ineq}). The most interesting part is the way in which Einstein treats the distribution of energies among the system as a collective state of a system made of several subsystems and, at the same time, as one possible macroscopic state of a system governed by a canonical distribution at the temperature~$T_{0}$. This device will be put to use in the 1910 work on critical fluctuations.~\cite{Einstein1910}

The physical interpretation of the constant~$\kappa=k_{\mathrm{B}}/2$ is obtained by Einstein in~\S~3 by evaluating, via his equipartition theorem, the kinetic energy of a mechanical system of~$n$ particles, and by relating the resulting expression to the one obtained by the kinetic theory for the ideal gas. He thus obtains an explicit estimate of $\kappa=6.5\cdot 10^{-17}\unit{erg\, K^{-1}}$, corresponding to~$k_{\mathrm{B}}=1.3\cdot 10^{-23}\unit{J\,K^{-1}}$. The discrepancy with modern values is due to the use of the value $N_{\mathrm{A}}=6.4\cdot 10^{23}\unit{mol^{-1}}$ for Avogadro's number, that Einstein found in O. E. Meyer's book.~\cite{Meyer}

In~\S~4, under the title ``General meaning of the constant~$\kappa$'' Einstein discusses the fluctuations of the energy in the canonical ensemble, deriving the relation between the specific heat and the amplitude of energy fluctuations as 
\begin{equation}\label{eq:FDT}
\average{E^{2}}-\average{E}^{2}=k_{\mathrm{B}}T^{2}\frac{d\average{E}}{d T},
\end{equation}
where $\average{\ldots}$ denotes the canonical average. Gibbs had obtained the same expression in~\cite[eq.~(205), p.~72]{Gibbs}, but pointed out almost immediately that these fluctuations were not observable. With $\epsilon$, $\epsilon_{p}$ and~$\epsilon_{q}$ the total, kinetic and potential energies respectively, and denoting averages by a bar, he states~\cite[p.~74f]{Gibbs}
\begin{quotation}
It follows that to human experience and observation with respect to such an ensemble as we are considering, or with respect to systems which may be regarded as taken at random from such an ensemble, when the number of degrees of freedom is of such order of magnitude as the number of molecules in the bodies subject to our observation and experiment, $\epsilon-\bar{\epsilon}$, $\epsilon_{p}-\bar{\epsilon}_{p}$, $\epsilon_{q}-\bar{\epsilon}_{q}$ would be in general vanishing quantities, since such experience would  not be wide enough to embrace the more considerable divergencies from the mean values, and such observation not nice enough to distinguish the ordinary divergencies. In other words, such ensembles would appear to human observation as ensembles of uniform energy, and in which the potential and kinetic energies (supposing that there were means of easing these quantities separately) had each separately uniform values.
\end{quotation}
Characteristically, Einstein instead goes over immediately to look for a system in which these fluctuations could be observed and he finds that the blackbody radiation could provide such a system. It is worth quoting his reasoning \cite[\S~5]{Einstein04}
\begin{quotation}
If the linear dimensions of a space filled with temperature radiation are very large in comparison with the wavelength corresponding to the maximum energy of the radiation at the temperature in question, then the mean energy fluctuation will obviously be very small in comparison with the mean radiation energy of that space. In contrast, if the radiation space is of the same order of magnitude as that wavelength, then the energy fluctuation will be of the same order of magnitude as the energy of the radiation of the radiation space.
\end{quotation}
Einstein pauses only one moment before proceeding to the application of his molecular theory of heat to black-body radiation~\cite[\S~5]{Einstein04}
\begin{quotation}
Of course, one can object that we are not permitted to assert that a radiation \textit{space} should be viewed as a \textit{system} of the kind we have assumed, not even if the applicability of the general molecular theory is conceded. Perhaps one would have to assume, for example, that the boundaries of the space vary with its electromagnetic state. However, these circumstances need not be considered, as we are dealing with orders of magnitude only.
\end{quotation}
Einstein can thus evaluate the size $\average{\epsilon^{2}}$ of the energy fluctuations $\epsilon=E-\average{E}$ from equation~(\ref{eq:FDT}) and from the Stefan-Boltzmann law
\begin{equation}
\average{E}=a\, v\, T^{4},
\end{equation}
where $a=7.06\cdot 10^{-15}\,\unit{erg\,cm^{-3}\,K^{-4}}$ is the radiation constant, $T$ is the absolute temperature, and~$v$ is the cavity volume. Then, the linear dimensions of a cavity for which $\average{\epsilon^{2}}\simeq \average{E}$ are given by
\begin{equation}
\sqrt[3]{v}=\frac{1}{T}\sqrt[3]{\frac{4 k_{\mathrm{B}}}{a}}=\frac{0.42}{T},
\end{equation}
which compares well (in order of magnitude) with the expression $\lambda_{\max}=0.293/T$ obtained from Planck's law (both lengths are expressed in cm, and $T$ is expressed in Kelvin).

However, in the following months, trying to explicitly apply his theory to that system, he will encounter a paradox, which he will brilliantly overcome by renouncing the classical picture of the emission and adsorption of light, based on Maxwell's equations, and by introducing the concept of the light quanta.~\cite{Einstein05a} But that is another story, which has already been told many times.
\section{Einstein and Gibbs}
\label{sec:EG}
One usually takes for granted that the research projects pursued by Einstein in these three papers, and by Gibbs in his 1902 book~\cite{Gibbs} were equivalent, and that the more mathematically refined argumentation contained in the latter made Einstein's approach redundant. A closer scrutiny shows however fundamental differences in their approaches, and makes Einstein's approach more attractive to present-day physicists. Gibbs program focuses in understanding the properties of \textit{ensembles} of mechanical systems, i.e., of systems whose dynamical equations are given, but whose initial conditions are only given in a probability distribution. He gives this discipline the name of ``statistical mechanics''. He stresses that its relevance goes beyond establishing a foundation of thermodynamics~\cite[Preface, p.~viii]{Gibbs}
\begin{quotation}
But although, as a matter of history, statistical mechanics
owes its origin to investigations in thermodynamics, it seems
eminently worthy of an independent development, both on
account of the elegance and simplicity of its principles, and
because it yields new results and places old truths in a new
light in departments quite outside of thermodynamics.
\end{quotation}
Indeed, statistical mechanics laws are \textit{more general} than those of thermodynamics~\cite[p.ix]{Gibbs}
\begin{quotation}
The laws of thermodynamics, as empirically determined,
express the approximate and probable behavior of systems of
a great number of particles, or, more precisely, they express
the laws of mechanics for such systems as they appear to
beings who have not the fineness of perception to enable
them to appreciate quantities of the order of magnitude of
those which relate to single particles, and who cannot repeat
their experiments often enough to obtain any but the most
probable results. The laws of statistical mechanics apply to
conservative systems of any number of degrees of freedom,
and are exact.
\end{quotation}
On the other hand, according to Gibbs, our ignorance of the basic constitution of material bodies make unreliable our inferences based on supposed models of matter, even when derived by the methods of statistical mechanics~\cite[p.ix-x]{Gibbs}
\begin{quotation}
In the present state of science, it seems hardly possible to frame a dynamic theory of molecular
action which shall embrace the phenomena of thermodynamics, of radiation, and of the electrical manifestations which accompany the union of atoms. [\dots]
Even if we confine our attention to the phenomena distinctively thermodynamic, we do not escape difficulties in as simple a matter as the number of degrees of freedom of a diatomic gas. It is well known that while theory would assign to the gas six degrees of freedom per molecule, in our experiments on specific heat we cannot account for more than five. Certainly, one is building on an insecure foundation, who rests his work on hypotheses concerning the constitution of matter.
\end{quotation}
Gibbs therefore attempts to reduce his goal to a purely mathematical treatment~\cite[p.~x]{Gibbs}
\begin{quotation}
Difficulties of this kind have deterred the author from attempting to explain the mysteries of nature, and have forced him to be contented with the more modest aim of deducing some of the more obvious propositions relating to the statistical branch of mechanics. Here, there can be no mistake in regard to the agreement of the hypotheses with the facts of nature, for nothing is assumed in that respect. The only error into which one can fall, is the want of agreement between the premises and the conclusions, and this, with care, one may hope, in the main, to avoid.
\end{quotation}
One can therefore only hope to establish \textit{analogies} between quantities which are defined within statistical mechanics, and those which are empirically encountered in thermodynamics~\cite[p.~x]{Gibbs}
\begin{quotation}
We meet with other quantities, in the development of the subject, which, when the number of degrees of freedom is very great, coincide sensibly with the modulus, and with the average index of probability, taken negatively, in a canonical ensemble, and which, therefore, may also be regarded as corresponding to temperature and entropy. 
\end{quotation}
The relations of the laws of statistical mechanics with thermodynamics is further discussed in~\cite[Ch.~XIV, p.~166]{Gibbs}
\begin{quotation}
A very little study of the statistical properties of conservative systems of a finite number of degrees of freedom is sufficient to make it appear, more or less distinctly, that the general laws of thermodynamics are the limit toward which the exact laws of such systems approximate, when their number of degrees of freedom is indefinitely increased. And the problem of finding the exact relations, as distinguished from the approximate, for systems of a great number of degrees of freedom, is practically the same as that of finding the relations which hold for any number of degrees of freedom, as distinguished from those which have been established on an empirical basis for systems of a great number of degrees of freedom.

The enunciation and proof of these exact laws, for systems of any finite number of degrees of freedom, has been a principal object of the preceding discussion. But it should be distinctly stated that, if the results obtained when the numbers of degrees of freedom are enormous coincide sensibly with the general laws of thermodynamics, however interesting and significant this coincidence may be, we are still far from having explained the phenomena of nature with respect to these laws. For, as compared with the case of nature, the systems which we have considered are of an ideal simplicity. [\dots] The phenomena of radiant heat, which certainly should not be neglected in any complete system of thermodynamics, and the electrical phenomena associated with the combination of atoms, seem to show that the hypothesis of systems of a finite number of degrees of freedom is inadequate for the explanation of the properties of bodies.
\end{quotation}

In Gibbs' approach, the probability distribution is a datum of the problem, while in Einstein's one it is one of the unknowns. The greatest difference is that Gibbs starts from the equal a priori probability postulate, while for Einstein what is important is to evaluate time averages and these are replaced by phase space averages through an ergodic hypothesis. Thus Gibbs is allowed to introduce the canonical distribution \textit{a priori}, as a particularly simple one, endowed with interesting properties, in particular because it factorizes when one considers the collection of two or more mechanically independent systems~\cite[Ch.~IV, p.~33]{Gibbs}
\begin{quotation}
The distribution [\dots] seems to represent the most simple case conceivable, since it has the property that when the system consists of parts with separate energies, the laws of the distribution in phase of the separate parts are of the same nature, a property which enormously simplifies the discussion, and is the foundation of extremely important relations to thermodynamics.
\end{quotation}
On the contrary, for Einstein, the canonical distribution is the distribution which describes the mechanical state of a system in contact with a thermal reservoir at a given temperature, while the ``simplest'' distribution is rather the microcanonical, which represents the state of an isolated system at equilibrium. And the former is derived from the latter.

Einstein's 1910 lecture notes on the Kinetic Theory of Heat at the University of Zurich show, in Navarro's words~\cite[\S6.2]{Navarro}, how his approach allowed him to proceed to
\begin{quotation}
the systematic application of statistical mechanics, once the canonical distribution is attained, to a large variety of fields. This is a sample list of the applications presented in the lecture notes: paramagnetism, Brownian motion, magnetic properties of solids, electron theory of metals, thermoelectricity, particle suspensions and viscosity. Gibbs invented, instead, a method whereby he could find no direct physical application other than the detection of the already mentioned thermodynamic analogies. Had Gibbs lived longer (he died the year after the publication of Elementary Principles) this might have changed. But, given his rigorous and extremely cautious attitude, any assumption on the issue is enormously risky.
\end{quotation}

Even more strikingly, in Einstein's hands, deviations from the expected behavior become a tool for the investigation of the microscopic dynamics. This difference in attitude was already highlighted above, in the discussion of energy fluctuations, but the clearest example is the 1905 paper on light emission and adsorption,~\cite{Einstein05a} where one notably reads
\begin{quotation}
This relation,\footnote{It is the relation now known as Jean's radiation law.} found as a condition for the dynamical equilibrium, not only fails to agree with the experiments, but also intimates that in our model a well-defined distribution of the energy between ether and matter is out of the question. [\dots] In the following, we shall treat the ``black-body radiation'' in connection with the experiments, without establishing it on any model of the production or propagation of the radiation.
\end{quotation}
Thus Einstein brackets the contemporary models of light adsorption and propagation, but maintains the statistical interpretation of entropy. He then evaluates the radiation entropy from the empirical distribution law and interprets it in terms of the statistical approach as describing the coexistence of point-like particles in a given volume (cf.~\cite{Norton}). This paper was soon followed by the equally bold application of Planck's radiation theory to the specific heats of solids~\cite{Einstein07}. 
\section{Concluding remarks}
\label{sec:Conclusion}
We presented Einstein's approach to statistical mechanics in contrast to the one taken by Gibbs. The results are equivalent since both are based on Boltzmann's contributions. Gibbs' starting point is the equal a~priori probability hypothesis in phase space that leads to the microcanonical probability density for an ensemble (of representative systems, according to Tolman~\cite{Tolman}). Einstein, on the other hand, starts by stating that what is important is the evaluation of time averages of appropriate quantities. These can be replaced by averages of the same quantities over an unknown density function over the phase space, with the help of an ergodic hypothesis.  Einstein introduces the assumption that the energy is the only conserved quantity to play the role of the ergodic hypothesis. Using this assumption and Lioville's theorem, Einstein shows that the unknown density function mentioned before must be constant on the energy shell, that is it must be the microcanonical distribution. From there, the interpretation of the canonical distribution is different: for Gibbs, it is the simplest distribution, which leads to describe as statistically independent systems which are physically independent, while for Einstein it is the distribution which describes the state of a system in contact with a reservoir. Thus the index of the canonical distribution (as defined by Gibbs) is \textit{analogous} to the temperature for Gibbs, but can be \textit{identified} with the temperature for Einstein. It is also interesting to remark that in several points Einstein states (without proof) that the distribution of energy values in the canonical ensemble is sharply peaked, and deduces from this some dubious inequalities for the probability density itself. Only in the 1904 paper he explicitly evaluates the size of fluctuations, obtaining a result already derived by Gibbs. Then, while Gibbs had stressed the non-observability of energy fluctuations in macroscopic systems (thus contributing to the ``rational foundation of thermodynamics''), Einstein points at the use of fluctuations as a tool for investigating microscopic dynamics (as he did, in particular, in~\cite{Einstein09}, where he hinted at the dual wave-particle nature of radiation by interpreting the two terms appearing the expression of energy fluctuations).

What interest can a present-day reader find in these papers? We think that they sketch a very neat road map for the introduction of the basic concepts of statistical mechanics, focusing on their heuristic value. One first focuses on isolated systems and identifies the microcanonical ensemble as the equilibrium distribution by means of the thermal equilibrium principle. For this step, Einstein's reasoning given above, based on the postulate of the absence of integrals of motion beyond the energy, is excellent. Then, one looks at a small part of such an isolated system, and one shows that the corresponding distribution is the canonical one. Finally, one identifies the mechanical expressions of temperature, infinitesimal heat and, by integration, of entropy. All these steps can be tersely traced by following, more or less closely, Einstein's path. At this point, the focus can be shifted to the evaluation of fluctuations, which allow on the one hand to recover the equivalence of ensembles for large enough systems and, by the same token, to identify situations in which the underlying molecular reality shows up in the behavior of macroscopic systems (like, e.g., in Brownian motion). This road map has been more or less followed by several modern textbooks on statistical mechanics, but we think that it would be fair to stress that it had first been sketched in the papers we described.

In any case, we will be satisfied if the present note encourages some colleagues to have a look at these papers, in which the first steps in the making of a giant are recorded. 
\section*{Acknowledgments}
LP was introduced to critical phenomena by Leo's lectures in the 1971 Varenna School, and RR fondly remembers Leo's course in the Escuela Mexicana de Física Estadística, which had a great influence on the Statistical Physics group at UNAM. Both authors dedicate this work to Leo's memory.

LP is grateful to Jeferson Arenzon for encouraging him to present his ideas on Einstein's 1902--04 works.

\end{document}